\documentstyle[twocolumn,prl,aps]{revtex}

\begin{document}
\draft
\title{Electron renormalization of sound interaction with two-level systems in
superconducting metglasses}
\author{E.V. Bezuglyi$^{1}$, A.L. Gaiduk$^{1}$, V.D. Fil$^{1}$,
W.L. Johnson$^{2}$, G. Bruls$^{3}$, B. L\"{u}thi$^{3}$, B. Wolf$^{3}$, and
S. Zherlitsyn$^{1,3}$}
\address{$^{1}$B.Verkin Institute for Low Temperature Physics and Engineering, 47\\
Lenin Avenue, Kharkov 310164, Ukraine}
\address{$^{2}$California Institute of Technology, Pasadena, CA 91125, USA}
\address{$^{3}$Physikalisches Institut, Universit\"{a}t Frankfurt, Robert
Mayerstr.2-4, 60054 Frankfurt, Germany }
\maketitle

\begin{abstract}
The crossing of temperature dependences of the sound velocity in normal and
superconducting state of metallic glasses points out unambiguously the
renormalization of the intensity of sound interaction with two-level systems
(TLS), caused by their coupling with electrons. We propose a simple scenario
which allows us to describe qualitatively the influence of the
superconducting transition on the sound velocity and attenuation in
superconducting metglasses and to make quantative estimations of
renormalization parameters.
\end{abstract}

\pacs{PACS numbers: 61.43.Fs, 62.65+k, 74.25.Ld}
\narrowtext

Acoustic measurements in superconducting metglasses Pd$_{30}$Zr$_{70}$
\cite{Neck1,Esq2}, Cu$_{30}$Zr$_{70}$ \cite{Esq3}, and
(Mo$_{1-x}$Ru$_{x}$)$_{0.8}$P$_{0.2}$ \cite{Licht4}, carried out about
a decade ago, have discovered, on the face of it, considerable contradictions
with the predictions of a well-known tunneling model (TM) \cite{Hunk5}. Below
we summarize the experimental facts which lead the authors of Refs. \cite
{Neck1,Esq2,Esq3,Licht4} to such a conclusion.

i) The sound velocity $v$ in superconducting (s) phase just below $T_{c}$ is
less than in normal (n) state. This effect was observed both in
low-frequency (LF) vibrating-reed measurements \cite{Neck1,Esq3,Licht4}, and
in high-frequency (HF) experiments \cite{Esq2}. According to the original TM,
the sound velocity should always increase below $T_{c}$ due to freezing out
of the negative relaxation contribution to $v$.

ii) The sound attenuation $\Gamma $ reveals an analogous anomaly: $\Gamma_s$
exceeds $\Gamma_{n}$ over some temperature interval, which is about of
$T/T_{c}\geq 0.8$ in HF measurements \cite{Esq2}, and extends up to
$T/T_{c}\geq 0.05$ in LF experiments \cite{Neck1,Esq3}, where $d\Gamma /dT$
has a large negative value. In contrast, TM predicts the attenuation to be
nearly independent of $T$ (with a small $d\Gamma /dT>0$) until the maximum
relaxation rate exceeds the sound frequency $\omega $. Thus, the attenuation
in LF experiments should be insensitive at all to superconducting
transition, whereas in HF measurements $\Gamma_{s}$ should either be
temperature-independent just below $T_{c}$, or decrease at low enough $T_c$.

iii) The slope of linear dependence $v(\ln T)$ in normal phase at HF is 4
times less than in superconducting state at $T\ll T_{c}$ \cite{Esq2},
whereas TM yields the ratio of 1:2.

iv) At least in HF experiments, the normal-state line $v_n(T)$ crosses the
superconducting dependence $v_s(T)$ at $T \ll T_c$ \cite{Esq2}. From the
point of view of TM, this is impossible in principle (see below).

v) Within a wide temperature range, the sound attenuation in the normal state
reveals a noticeable temperature dependence with $d\Gamma/dT > 0$, which
significantly exceeds the changes calculated on the basis of TM.

vi) In contrast to TM predictions, $dv/dT<0$ in LF experiments \cite
{Neck1,Esq3} in normal state at $T<1$K, where the phonon relaxation can be
neglected. On the other hand, in HF experiments \cite{Esq2} $dv/dT>0$ within
the same temperature interval.

It was supposed in Refs. \cite{Neck1,Esq2} that all (or most of all)
deviations from TM are related to the electron renormalization of the
constant $C=\overline{p}\gamma ^{2}/\rho v^{2}$ (where $\overline{p}$ is the
TLS density of states, $\gamma $ is the deformation potential, $\rho $ is
the mass density), which determines the scale of changes of $v$ and $\Gamma$
in the presence of TLS. However, it was not proposed any consistent scheme
to trace, even qualitatively, the genesis of peculiarities of
superconducting metglasses mentioned above. Also, there were no attempts to
estimate either the value of renormalization of $C$, or its possible energy
dependence.

In this work, we have investigated low-temperature HF acoustic properties of
superconducting amorphous alloy
Zr$_{41.2}$Ti$_{13.8}$Cu$_{12.5}$Ni$_{10}$Be$_{22.5}$ \cite{John6} and found
similar discrepances with the TM. Following the suggestion of \cite{Neck1,Esq2}
we propose a simple scenario, which allows us to explain most of the puzzling
experimental facts within the framework of TM by introducing an
energy-dependent electron renormalization of $C$. It is supposed to be
significant up to some characteristic energy $E_{k}$ and to be suppressed in
the superconducting state.

The alloy under investigation has a high resistance for crystallization in
the state of supercooled melt and remains amorphous at extremely low cooling
rate ($<$ 10 K/s) \cite{John6}. This makes it possible to obtain bulk
homogeneous samples, which suit perfectly the acoustic measurements.
Ultrasonic experimental technique is described elsewhere \cite {Luthi}.

The experimental dependences $\delta v(\ln T)/v$ of the transverse sound for
normal and superconducting phases are presented in Figure \ref{fig1} by
filled and open symbols, correspondingly. The normal-state measurements were
carried out in the magnetic field B = 2.5 T. In accordance with TM, the
plots $v(\ln T)$ represent almost straight lines both in the normal and in
the deep superconducting state (here and below all the experimental data are
normalized on the slope value in the deep superconducting phase,
$C=2.85\cdot 10^{-5}$). The growth of $v$ below $T_{c}$ reflects the freezing
out of its relaxation component and agrees with the TM conception. There are
also obvious deviations from the TM: the ratio of slopes in the normal and
the superconducting phases differs from its canonical value 1:2 and is close
to 1:4, and the curves $v(T)$ for both phases intersect at some
frequency-dependent temperatures $T_{cr}$. Such effects have been observed
before in Pd$_{30}$Zr$_{70}$ [2].

In general, the ``anomalous'' slope ratio can be explained by the energy
dependence of $C$, which results in different contributions into the
resonant and relaxation components of $v$. However, within such a simple
approach, the crossing remains unaccountable, because it would mean the
change of sign of the relaxation contribution below $T_{cr}$, that
contradicts the physical sense of a relaxation process.

The only possibility to explain the crossing is to assume the growth of $C$
in the superconducting phase, as the result of the suppression of the electron
renormalization at energies less than the superconducting gap. To validate
this assumption, we use the expression for the resonant contribution of TLS
into the sound velocity with account of energy dependence of $C$ \cite{Piche}:
\begin{equation}
\left( {\frac{\delta v(T)}{v}}\right)_{res}=P\int_{0}^{\infty }{
\frac{C(\omega ^{\prime })\omega ^{\prime }\tanh (\omega ^{\prime }/2T)}
{\omega^{2}-\omega ^{^{\prime }2}}}d\omega ^{\prime }
\end{equation}
Here and below the energy, the temperature and the frequency are expressed
in the same units.

In a simple case of energy-independent $C$, to avoid the logarithmic
divergence of Eq. (1) at the upper limit, this formula is conventionally
used to obtain an expression for the relative variations of $v$ counted off
its value at some arbitrary reference temperature $T_{0}$:
\begin{equation}
\left( \delta v(T)/v\right)_{res}=C\ln {(T/T}_{0}),\qquad T>\omega
\end{equation}

Obviously, if the value of $C$ varies with energy and/or temperature, Eq.
(2) is inapplicable even for qualitative estimates, since the reference
value of $\delta v/v$ may also change with $C$. To account correctly for the
changes of $C$, it is necessary to analyze the complete integral of Eq. (1)
by introducing a cut-off energy $E_{m}$ (say, of the order of melting
temperature). It describes a real negative contribution of resonant
transitions in the TLS system into the sound velocity. Within the
logarithmic accuracy, in the case of $C=$ const Eq. (1) can be approximated
by the following piece-linear dependence (see Figure \ref{fig2}, line 1):
\begin{equation}
\left( {\frac{\delta v(T)}{Cv}}\right)_{res}=\left\{
\begin{array}{cc}
\ln (\omega /E_{m}) & T<\omega \\
\ln (T/E_{m}) & T>\omega
\end{array}
\right.
\end{equation}
Here we neglect small variations in $v$ at $T\lesssim \omega $: a quadratic
fall near $T=0$ and a shallow minimum at $\omega =2.2T$ \cite{Gold7}.

The similar dependence for ``renormalized'' constant $C^{\prime }=C(1-R)<C$,
also plotted in Figure \ref{fig2} (line 2), appears to be located above the
first one for $T<E_{k}$. Here we use the simplest step-like model of
renormalization: $R=R_{0}>0$ for $E<E_{k}$, and $R=0$ in opposite case.
Thus, the sound velocity increases when $C^{\prime}$ reduces
(at $T<E_{k}$), and vice versa; this basic conclusion can not be derived from
Eq. (2), where the ``dc component is lost''.

A numerical calculation shows, that the relaxation contribution $\left(
{\delta v(T)/v}\right)_{rel}$ at $T\lesssim 1K$, where only the electron
relaxation is significant, is also well approximated by piece-linear
dependence
\begin{equation}
\left( {\frac{\delta v(T)}{Cv}}\right)_{rel}={\frac{1-R}{2}}\left\{
\begin{array}{cc}
0 & T<\omega /\eta ^{2} \\
\ln (\omega /T\eta ^{2}) & T>\omega /\eta ^{2}
\end{array}
\right.
\end{equation}
with a dimensionless parameter $\eta \lesssim 1$ of coupling between the
electrons and TLS \cite{Hunk5}. The resulting $\delta v/v$ plot (line 3)
shows the velocity variation in normal state. It is clear, that the
intersection of lines 3 and 1 (the latter represents a sketch of the
velocity change in the deep superconducting state) is possible only under
the condition of $C^{\prime} < C$ in normal state.

Now we can explain qualitatively the behavior of $v$ and $\Gamma $ at the
superconducting transition. Below $T_{c}$ the electron renormalization
reduces rapidly, and the effective $C^{\prime}$ grows, providing the
decrease in $v$ and the increase in $\Gamma$. However, a competitive effect
arises simultaneously: the time $\tau $ of relaxation of TLS on electrons
grows and, therefore, changes $v$ and $\Gamma $ into the opposite direction.
Thus, if the phonon relaxation predominates near $T_{c}$, the effective
$\tau$ changes weakly, and the sound velocity will decrease
(correspondingly, $\Gamma $ will increase) below $T_{c}$, as it was observed
in \cite{Neck1,Esq2,Esq3,Licht4}. If, on the contrary, the electron
relaxation prevails (for materials with lower $T_{c}$ like our system), the
changes of $v$ and $\Gamma $ near $T_{c}$ may have any sign, depending upon
the relation between $T_{c}$ and $E_{k}$.

Following our approach, the observed slope ratio of $v(\ln T)$ plots might
be treated as the decrease of $C^{\prime}$ down to 0.5$C$ ($R_{0}=0.5$)
at $E<E_{k}$. Making use of known values of $T_{cr}$ (see Figure \ref{fig1})
and the elementary geometry of Figure \ref{fig2}, we obtain the value of
$E_{k}=$ 0.2 K, which, however, seems to be too low to reduce sufficiently
the TM ratio of slopes in normal state at $T\sim $ $1$K and in deep
superconducting state. In fact, these values of $R_{0}$ and $E_{k}$ are to
be interpreted as tentative estimates for a numerical analysis of Eq. (1),
implemented by us to compare quantitatively the proposed model with the
experimental data.

For this purpose, we need first the value of $\eta $ which can be found from
the temperature dependence of sound attenuation in superconducting state,
$\Gamma_{s}(T)/\Gamma_{n}(T_{c})$, plotted in Figure \ref{fig3}, under
assumption of validity of conventional BCS theory \cite{foot}. According to
TM and the theoretical dependence of $\tau $ in the superconducting phase
\cite{Black9}, the low-temperature ($T\ll \Delta $) behavior of attenuation is
described by the exponential dependence $\Gamma_{s}(T)/T\Gamma
_{n}(T_{c})\sim \exp (-\Delta (0)/T)$, confirmed by the experimental data
plotted in the inset of Figure \ref{fig3}. Note, that the slope of the
linear part of this curve yields the BCS ratio of $\Delta (0)/T_{c}=1.73\pm
0.1$ ($T_{c}=0.9K$). A numerical calculation of $\Gamma_{s}(T)$ with the
fitting value $\eta =0.84$ describes satisfactorily the evolution of
attenuation over the actual temperature region below $T_{c}$ as shown by
solid line in Fig.3.

In our numerical calculations, we have used the model of a more realistic
smoothed energy dependence of $C^{\prime}$:
\begin{equation}
C^{\prime}(E)/C=1-R_{0}\left( 1-\tanh ^{2}(E/E_{k})\right)
\end{equation}

The two fitting parameters $0<R_{0}<1$ and $E_{k}$ were found to be
determined by the slopes of $\delta v(\ln T)$ in the normal state and by the
crossing temperatures $T_{cr}$ at two frequences. In order to extend our
model over the superconducting state, we have accounted for the freezing out
of normal electron excitations with additional multiplying of the
renormalization parameter $R_{0}$ by the freezing factor $2f(\Delta /T)$ (
$f(x)$ is the Fermi function) at energies less than the superconducting gap
$2\Delta $. Of course, this approximation ignores the energy dispersion of
the electron density of states in the superconductor, but, nevertheless, it
reproduces perfectly the experimental velocity dependence, as it is obvious
from Figure \ref{fig1}. The calculated velocity dependences, with account of
relaxation contribution, deviate from experimental data at most of noise
level within a narrow interval of $E_{k}=1.75\pm 0.25K$ and $R_0=0.27\pm 0.01
$. Note that, according to Eqs. (1), (3), $(\delta v/v)_{res}$ does not
depend on frequency at $T\gg \omega $ . This allowed us to compare the
results at different frequences for a given $C$ by aligning the
low-temperature parts of the experimental curves, where only the resonant
component is present.

The changes of relaxation and resonant contributions into the sound velocity
almost compensate each other near $T_{c}$ for our $R_{0}$ and $E_{k}$, with
the creation of a shallow minimum (of the noise level). Apparently, this
causes the mismatch between critical temperatures, found in \cite{Gai8} by
the magnetic (0.9 K) and acoustic (0.85 K) measurements (see also Figure 1).
For larger $T_{c}\gtrsim 2\div 3$K, the relaxation component of $v$,
determined mainly by the interaction of TLS with phonons, changes more
slowly, and the sound velocity below $T_{c}$ should decrease much stronger.

As regards the relaxation contribution to sound attenuation, in the normal
state the system ``scans'' the relation of Eq. (5), therefore $\Gamma_{n}(T)$
decreases with temperature, and the TM plateau is absent in fact. The
evolution of $\Gamma_{s}(T)$ at the superconducting transition is
determined first by the suppression of renormalization, caused by freezing
out of normal excitations, that results in the growth of attenuation
($\sim\Delta^{2}(T)$). Such a behavior agrees qualitatively with the
experiments of Refs.\ \cite{Neck1,Esq2,Esq3,Licht4}. However, in our case
the experimental changes of attenuation are noticeably less than calculated
ones, as it is shown in Figure \ref{fig4}. This is obviously a consequence
of roughness of our model, and we shall outline a way to remove this
imperfection.

Following the results of Ref. \cite{Vlad10}, the renormalization of constant
$C$ seems to be dependent not only on the energy $E$, but also on the
splitting parameter $u=\Delta_{0}/E$, where $\Delta_{0}$ is the tunnel
splitting in symmetric double-well potential. Both of them determine the
relaxation parameter $\omega \tau \sim (\omega /u^{2}E)\tanh (E/2T)$, which
is of the order of $\omega \tau \sim (\omega /u^{2}T)$ for all actual
energies $E\leq T$. The relaxation contribution into the sound velocity is
formed by TLS with $\omega \tau \ll 1$, or, equivalently, $u^{2}\gg \omega
/T\sim 10^{-2}$, that corresponds to nearly symmetric TLS. In contrary, the
attenuation is contributed by asymmetric TLS with $\omega \tau \sim 1$, or
$u^{2}\sim \omega /T$. It was shown in \cite{Vlad10} that the influence of
electrons on TLS is mainly significant in the symmetric case. Thus, the
summary of our analysis can be treated as the difference between the
effective constants $C_{v}^{\prime}$ and $C_{\Gamma }^{\prime}$,
describing the renormalization of symmetric and asymmetric TLS,
correspondingly. After the averaging over $u$, the constant $C_{v}^{\prime}$
is well described by the model expression of Eq. (5) with fitting
parameters found above. Apparently, $C_{\Gamma }^{\prime}$ can be
described by an analogous formula, but with much smaller renormalization
parameter $R_{0}\sim 0.05$.

In that way, the proposed scenario allows us to explain qualitatively all
discrepancies between the real experimental pattern and the original TM,
mentioned in the preamble, except the item vi). Indeed, in our model the
derivative $dv/dT$ is always positive at $T<1K$ (though less than in simple
TM). At present time, within a scope of TM or its modifications, there are
no mechanism to explain the change of sign of $dv/dT$ upon decreasing of
frequency, when the condition $\omega \tau \ll 1$ still holds. This problem
needs further experimental investigations over the intermediate frequency
range.

In summary, the crossing of the temperature dependences of sound velocity in
normal and superconducting states of superconducting metglasses at $T\ll
\Delta $ unambiguously confirms the existence of electron renormalization of
the constant $C=\overline{p}\gamma ^{2}/\rho v^{2}$ towards smaller values.
An assumption about the energy-dependent renormalization concentrated within
a low-energy region allows us to describe perfectly the behavior of sound
velocity in normal and superconducting phases by only two fitting parameters
within the framework of TM.

 This research was partially supported by Ukrainian State Foundation for
Fundamental Research (Grant No 2.4/153) and the Deutsche
Forschungsgemeinschaft via SFB 252. W.L.J. wishes to acknowlege the U.S. Dept.
of Energy for support under Grant No. DE-FG03-86ER45242. S.Z. would like to
thank Alexander von Humboldt-Foundation for support.

\begin{figure}[tbp]
\caption{Experimental (symbols) and calculated (solid curves) temperature
dependences of transverse sound velocity in
Zr$_{41.2}$Ti$_{13.8}$Cu$_{12.5}$Ni$_{10}$Be$_{22.5}$ for two frequencies:
circles - 62 MHz, triangles - 186 MHz. Open symbols represent the result for
the superconducting state, filled symbols -for the normal state,
$C = 2.85*10^{-5}$. Arrows indicate the critical temperature of the
superconducting transition $T_c$ found from magnetic measurements [11] and
the crossing temperatures $T_{cr}$. }
\label{fig1}
\end{figure}

\begin{figure}[tbp]
\caption{Diagram of the temperature dependences of the sound velocity in
superconducting metglass: 1 - $(\delta v/v)_{res}$ in the superconducting
phase, 2 - $(\delta v/v)_{res}$ in the normal phase, 3 - the total $\delta
v/v$ in the normal phase. }
\label{fig2}
\end{figure}

\begin{figure}[tbp]
\caption{Temperature dependence of sound attenuation in superconducting
state of Zr$_{41.2}$Ti$_{13.8}$Cu$_{12.5}$Ni$_{10}$Be$_{22.5}$: circles -
experiment at 62 MHz, solid curve - calculation with $\eta$ = 0.84. Inset:
to determination of the superconducting energy gap.}
\label{fig3}
\end{figure}

\begin{figure}[tbp]
\caption{Comparison of the sound attenuation at 54 MHz with model
calculation: open and solid symbols - experimental dependences in normal and
superconducting phases, correspondingly; lines - the related calculation for
$E_k=1.75$ K, $R_0 = 0.27$. Note that the vertical scale is strongly enlarged.
}
\label{fig4}
\end{figure}


\begin{references}
\bibitem{foot}  The latter, generally, was not obvious, since the critical
temperature of studied alloy found from AC magnetic measurements does not
coincide with a sharp bend on $v(T)$ \cite{Gai8}, that may be attributed to
the effect of paramagnetic depairing.
\end{references}
\end{document}